\documentclass{article}
\usepackage{authblk}
\usepackage[stable]{footmisc}

\usepackage{hyperref}
\usepackage{amsmath}
\usepackage{amsfonts}
\usepackage{amssymb}

\usepackage{geometry}
\usepackage{float}
\usepackage{tikz}
\usepackage{pgfplots}
\pgfplotsset{compat=1.11}

\tikzset{viewport/.style 2 args={
    x={({cos(-#1)*1cm},{sin(-#1)*sin(#2)*1cm})},
    y={({-sin(-#1)*1cm},{cos(-#1)*sin(#2)*1cm})},
    z={(0,{cos(#2)*1cm})}
}}

\newcommand{\ToXYZ}[2]{
    {sin(#1)*cos(#2)}, 
    {cos(#1)*cos(#2)}, 
    {sin(#2)}          
}

\newcommand{\beq}{\begin{equation}}
\newcommand{\eeq}{\end{equation}}
\newcommand{\ba}{\begin{array}}
\newcommand{\ea}{\end{array}}
\newcommand{\beqa}{\begin{eqnarray}}
\newcommand{\eeqa}{\end{eqnarray}}

\title{Potentials on the conformally compactified Minkowski spacetime and their application to quark  deconfinement}
\author[1]{M. Kirchbach}
\author[2]{J. A. Vallejo}
\affil[1]{Instituto de Física\\ 
Universidad Autónoma de San Luis Potosí\\ 
Av. Chapultepec 1570, San Luis Potosí, SLP 78295, Mexico\\
\texttt{mariana@ifisica.uaslp.mx}}
\affil[2]{Departamento de Matemáticas Fundamentales\\ 
Universidad Nacional de Educación a Distancia\\ 
C. Juan del Rosal 10, CP 28040, Madrid, Spain\\
\texttt{jvallejo@mat.uned.es}}

\begin{document}

\maketitle

\def\RotationX{0}
\def\RotationY{0}

%
%
%
%
%
%
%


\begin{abstract}
We study a class of conformal metric deformations in the quasi-radial coordinate parameterizing the 3-sphere in  the conformally compactified Minkowski spacetime $S^1\times S^3$. Prior to reduction of the associated Laplace-Beltrami operators to a Schr\"odinger form, a corresponding class of exactly solvable potentials (each one containing a scalar and a gradient term) is found. In particular, the scalar piece of these potentials can be exactly or quasi-exactly solvable, and among them we find the finite range confining trigonometric potentials of P\"oschl-Teller, Scarf and Rosen-Morse. As an application of the results developed in the paper, the large compactification radius limit of the interaction described by some of these potentials is studied, and this regime is shown to be relevant to a quantum mechanical quark deconfinement  mechanism.
\end{abstract}


\section{Introduction}
The conformally compactified Minkowski spacetime, $M^c\simeq S^1\times S^3$,  is known to be topologically equivalent to the  conformal boundary of $AdS_5 $ \cite{Schottenloher}. When considered as the configuration space of quarks and gluons, an \emph{ansatz} inspired by gauge--gravity duality conjecture, it only allows for neutral color-electric charge $2n$ quark--anti-quark systems on it.  Thus, $M^c$ is well suited for the description of color confinement in terms of color--electric charge dipoles, such as  mesons, which can be described by means of the conformal wave equation on this closed manifold \cite{KTV}. An appropriate potential can then be added to that equation, either coming from a solution to a certain Poisson equation on $S^3$, or through an appropriate conformal metric deformation, thus giving rise to interacting color--electric $2-$charge systems.
Such procedure leads to a description of color confinement within the framework of quantum potential models on the $AdS_5$ boundary \cite{KTV}, an approach complementary to that of QCD, based on gauge fields.

To be more precise, the  solution to the Poisson equation on $S^3$ for a color--electric charge dipole source has been shown in \cite{KC2016} to be given by the trigonometric Rosen-Morse potential (modulo an additive constant).
In \cite{KTV}, a potential of the same type has been shown to be the byproduct of a particular kind of conformal transformation of the $M^c$ metric (we will also use the term `conformal deformation' in the sequel). Therefore, the trigonometric Rosen-Morse potential makes its appearance through two mechanisms of different nature. 
While the first mechanism needs the existence of a color--electric charge dipole as a source, the second one has a purely geometrical origin, tied to certain features of $M^c$. This peculiarity suggests that the trigonometric Rosen-Morse potential can be a useful tool in the description be it of  quark--antiquark systems, as are the majority of the mesons, be it of  three quark systems, as are the baryons, when treated in the quark--anti-symmetric diquark approximation, or be it  of multi-quark systems such as the more recently detected  tetra-quarks.  This is documented  by its successful  employment  by several authors in the analysis of a broad range of hadron data including their spectra \cite{KC2016}, \cite{Ahmed}, \cite{AbuShady},  electromagnetic form factors \cite{MK_Formfactors}, dressing of the gluon propagator \cite{Marco},  and the transition of a charmonium quantum gas to a Bose-Einstein condensate \cite{AramDavid}. Also, let us remark that the relation between properties of QCD  and color di-electrics has been studied among others in \cite{Pirner}.

Our main goal in the present work is to construct, through arbitrary  $M^c$ metric deformations, a  new family of scalar potentials whose exact ground state solutions are determined by the corresponding conformal Laplacians. Among them one finds the already mentioned trigonometric Rosen-Morse potential,
the  (well known from super-symmetric quantum mechanics) exactly solvable trigonometric P\"oschl-Teller and Scarf potentials, and other potentials which are only quasi-exactly solvable. We refer to all of them as `induced' or `mean field' scalar potentials. Furthermore, we explore importance of the compactification radius in the physics of confinement, for the particular case of the trigonometric Rosen-Morse potential interpreted as a color-electric charge dipole potential. We show that for large compactification radii, as they can occur at high temperatures, the color--electric charge dipole potential can collapse to one of Coulomb-type, thus locally enabling the formation of multi-$q\bar q$ systems of Rydberg-atom type.  In this way, at some large compactification radii, the Rydberg  quark, occupying an orbital with very high principal quantum number, can become loosely bound and be knocked out in an inelastic scattering event, thus becoming accessible to observation.
Therefore, we address from first principles the fundamental question of the possible existence of a deconfined phase at very high temperatures.

The text is structured as follows. In the next section we briefly discuss the importance of the compactified Minkowski spacetime $M^c$ in strong interactions, and the properties of the conformal wave operator on it. In Section 3, to make our exposition more or less self-contained, we recall the
derivation  of the solution to the Poisson equation on $S^3$ with  a color-electric charge dipole source.  Section 4 is devoted to the construction of a
master formula describing a  new family  of scalar solvable potentials induced  by metric deformations on $M^c$, as well as to their exact ground state solutions. As examples, we present the three well known exactly solvable trigonometric potentials of Rosen-Morse, P\"oschl-Teller and Scarf, as well as a further example of a new, only quasi-exactly solvable potential.  In Section 5  we consider, as an application of the ideas previously presented,  a possible scenario for the formation of Rydberg atoms in the compactified Minkowski spacetime, in the limit of large compactification radii, and their ionization, leading to a feasible quark deconfinement mechanism. Also,
we explain how to reconcile the masslessness of the conformal wave equation on $M^c$ with the `reduced' mass parameter appearing in its reduction to a Schr\"odinger equation. The text finishes with some conclusions and outlook on further work.

\section{The conformally compactified Minkowski spacetime $M^c\simeq S^1\times S^3$}
Understanding the reasons for the non-observability of free quarks, a phenomenon known as `color confinement', is one of the millennium puzzles in contemporary particle physics, both theoretical and experimental. A proposal in that regard has been presented in 
\cite{KTV}, where it has been attributed  to the peculiarity of the internal space of hadrons, assumed to be a compact space on which, as a consequence of Stokes theorem, no free charges can occur. 
More specifically, that space has been suggested to be  the compactified conformally symmetric Minkowski spacetime, $M^c$, whose  topology $S^1\times S^3$, coincides with that of the $AdS_5$ boundary. The above framework allowed for the  construction, in \cite{KTV}, of the non-Euclidean quantum mechanical limit of QCD in the infrared regime, by  means of the conformal wave equation on the compactified $AdS_5$ boundary. Conformal symmetry furthermore  governs QCD in the ultraviolet regime  due to the vanishing of the strong coupling, and in this setting a quantum mechanical scheme (this time an Euclidean one)  has also been developed, on the basis of the Galilean conformal mechanics, and has been applied to the description of a broad range of QCD processes \cite{Brodski}. The common denominator of the two schemes is provided by the $AdS_5$ boundary, i.e. the $AdS_5$ null-ray cone, whose intersections with properly chosen  hyper-planes define the topologies of the internal space of the QCD degrees of freedom in the infrared, $S^1\times S^3$, 
and in the ultraviolet, where it becomes singular. This is in accord with the gauge-gravity duality principle, which suggests a conformally invariant strongly coupled field theory, assumed to be  QCD in the infrared regime,
to reside at the boundary of $AdS_5$,  an idea supported by the detected  walking of the strong coupling $\alpha_s$ to a fixed value, $\alpha_s/\pi{\longrightarrow} 1$, when $Q^2\to 0$ \cite{Deur}. The topologies of the  two aforementioned internal spaces are necessarily quite different indeed, as they  reflect the different quark-gluon dynamics characterizing the two extreme regimes of QCD.

\subsection{Conformal wave equation on $M^c$}

Parameterizing $S^3$ by the two polar angles $\chi,\theta \in [0,\pi]$, and the azimuthal one $\varphi\in [0,2\pi )$, the $M^c$ line element can be written as
\begin{equation}
{\mathrm d}s^2={\mathrm d}\tau^2 -{\mathrm d}\chi^2 +\sin^2\chi \left({\mathrm d}\theta^2 +\sin^2\theta {\mathrm d}\varphi^2\right).
\label{S3_mtrc}
\end{equation}
The corresponding conformal wave  operator for this metric  $g_{S^1\times S^3}$, is then \cite{BirrelDavis}
\begin{equation}
  \Box_{g_{S^1\times S^3}}(\tau,\chi,\theta,\varphi) =
    \frac{\partial ^2}{\partial \tau^2}- \Delta_{g_{S^3}}(\chi,\theta,\varphi)  +\frac{1}{6}
    \mbox{Scal} _{g_{S^1\times S^3}}  \,, 
\label{Canz1}
\end{equation}
where $\tau$ is the angle parameterizing the circle $S^1$.   
Here,  $\Delta_{ S^3}(\chi,\theta,\varphi)$   is the Laplace-Beltrami operator on $S^3$,  
\begin{eqnarray}
-\Delta_{S^3}(\chi,\theta,\varphi)&=& -\frac{\partial^2}{\partial \chi^2} -2\cot\chi \frac{\partial }{\partial \chi}+\frac{{\mathbf L}^2(\theta,\varphi)}{\sin^2\chi},
\end{eqnarray}
with  Scal$_{g_{S^1\times S^3}}$ the scalar  curvature of $S^1\times S^3$, given by
\begin{equation}
\mathrm{Scal}_{g_{S^1\times S^3}}=6,
\label{Sclr_curv_sphere}
\end{equation}
and ${\mathbf L}^2(\theta,\varphi)$ is the ordinary  squared
orbital angular momentum operator  on the 2-sphere, whose associated eigenvalue problem is
\begin{equation}
{\mathbf L}^2(\theta,\varphi)Y_\ell^m(\theta,\varphi)=\ell (\ell+1)Y_\ell^m(\theta,\varphi)\,.
\label{Lcuadrado}
\end{equation}
Assuming a factorization of the eigenfunctions of the operator (\ref{Canz1}) in the form
\begin{equation}
u_{g_{S^1\times S^3}}=e^{i(K+1)\tau} Y_{K\ell m} (\chi,\theta,\varphi)\,,
\label{fctrz_wf}
\end{equation}
the  conformal wave equation on $S^1\times S^3$ reads,
\begin{eqnarray}
\Box_{g_{S^1\times S^3}}(\tau,\chi,\theta,\varphi)e^{i\tau (K+1)}Y_{K\ell m}(\chi,\theta,\varphi)&=&0,
\label{cfrm_weq}
\end{eqnarray}
where  $Y_{K\ell m} (\chi,\theta,\varphi)$ are the well known ultra-spherical harmonics \cite{4harm}, defined by
\begin{eqnarray}
Y_{K\ell m}(\chi, \theta,\varphi) =S_{K\ell} (\chi) Y_{\ell }^m(\theta, \varphi),\mkern9mu
S_{K\ell}(\chi)=\sin ^{\ell }\chi G_{n}^{\ell +1}(\cos \chi),\mkern9mu n=K-\ell,
\label{Gegenb}
\end{eqnarray}
where $G_n^\alpha(\cos\chi)$ denote the  Gegenbauer polynomials of degree $n$ and parameter $\alpha$. Notice that $(n+\ell)$ stays constant and equals the value of the $K$ parameter in (\ref{fctrz_wf}).  As  we shall see in the subsequent section, this constant represents the  value of the four-dimensional angular momentum on $M^c$.

\subsection{The conformal wave operator in terms of the Casimir of $so(4)$}

Recalling the relationship between $\Delta_{S^3}(\chi,\theta,\varphi)$ and the squared four-dimensional angular momentum operator, ${\mathcal K}^2(\chi,\theta,\varphi)$ (the Casimir invariant of the $so(4)$ Lie algebra),
\begin{eqnarray}
-\Delta_{S^3}(\chi,\theta,\varphi)&=&{\mathcal K}^2(\chi, \theta, \varphi),
\label{Dlt_s3}
\end{eqnarray}
the conformal wave operator (\ref{Canz1}) can be expressed as
\begin{equation}
\Box_{g_{S^1\times S^3}}(\tau, \chi,\theta,\varphi)=\frac{\partial ^2}{\partial \tau^2} +{\mathcal K}^2(\chi,\theta,\varphi)+\frac{1}{6} \mbox{Scal} _{g_{S^1\times S^3}}\,,
\label{new_box_oprtr}
\end{equation}
a representation which will prove to be useful in what follows.

The massless equation defined by (\ref{Canz1}), (\ref{cfrm_weq}) now becomes,
\begin{equation}
\left[ -(K+1)^2 +{\mathcal K}^2(\chi,\theta,\varphi)  +1 \right]
Y_{K\ell m}(\chi,\theta,\varphi )=0\,,
\label{free_on_S3}
\end{equation}
which is the equation for a massless scalar particle in free stationary motion on $M^c$. So, in view of (\ref{Dlt_s3}), the $K$ label in the ultra-spherical harmonics 
acquires meaning of four-dimensional angular momentum value, as mentioned.

By separating the angular variables $(\theta, \varphi) $ from the `quasi-radial' variable $\chi$ as in \eqref{Gegenb}, equation (\ref{free_on_S3}) transforms into the following eigenvalue problem, 
\begin{eqnarray}
\left( -\Delta_{S^3}(\chi) +1 \right)S_{K\ell}(\chi)
&=& \left( {\mathcal K}^2(\chi) +1 \right)S_{K\ell}(\chi)
= (K+1)^2  S_{K\ell}(\chi),
\label{1d_Schr_ty}
\end{eqnarray}
with
\begin{eqnarray}
-\Delta_{S^3}(\chi)&=&{\mathcal K}^2(\chi)=-\frac{\partial ^2}{\partial \chi^2 } -2\cot\chi\frac{\partial }{\partial \chi} +\frac{\ell(\ell +1)}{\sin^2\chi}.
\label{rdcd_dltS3}
\end{eqnarray}
Here, $\Delta_{S^3}(\chi)$ and ${\mathcal K}^2(\chi)$ are, respectively, the $S^3$ Laplacian and the squared four-dimensional angular momentum operators reduced to the quasi-radial variable, in which case  
$\ell $ denotes a parameter  referring  to the second quantum number in the  labeling of its eigenfunctions, $S_{K\ell}(\chi)$, and no longer to the degree of the spherical harmonics $Y_\ell^ m(\theta, \varphi)$.

\subsection{Reduction to a one--dimensional Schr\"odinger equation}

Through the replacement
\begin{equation}
S_{K\ell}(\chi)=\frac{{\mathcal U}_{K\ell}(\chi)}{\sin \chi }\,,
\label{1st_chng}
\end{equation}
and the similarity transformation of the operator $\left( -\Delta_{S^3}(\chi) +1 \right)$ given by
\begin{equation}
\sin\chi \left( -\Delta_{S^3}(\chi) +1 \right)
\sin^{-1} \chi \left[ \sin\chi S_{K\ell}(\chi)\right] =
 (K+1)^2 \left[\sin \chi S_{K\ell}(\chi)\right]\,,
\label{smlrt_trnsfm}
\end{equation} 
the first-order derivative in  (\ref{rdcd_dltS3}) can be removed, and  (\ref{1d_Schr_ty}) now adopts the form of a Schr\"odinger--type equation,
\begin{equation}
\left(-\frac{\partial ^2}{\partial \chi^2} +\frac{\ell (\ell +1}{\sin^2 \chi} \right){\mathcal U}_{K\ell}(\chi)=\left( K+1\right)^2{\mathcal U}_{K\ell }(\chi),\quad \ell =0,1,2,..., K,
\label{degeneracy}
\end{equation}
where the centrifugal term in $\Delta_{S^3}(\chi)$ has now the role of a 
one-dimensional centrifugal potential.
Furthermore, this expression shows that the energy of a particle moving within the $S^3$ centrifugal potential does not depend on the $\ell$ value of the three dimensional angular momentum, but rather it
is $(K+1)^2$-fold degenerate, as it should be in view of the $SO(4)$ symmetry of the problem presented in the previous subsection.

In the remainder of the paper, we will explore the procedures for, and consequences of, the inclusion of potentials into (\ref{free_on_S3}), but before closing this section, we would like to remark 
an intriguing  approach to masslessness in $O(2,4)$ gauge theories,
 suggested by \cite{Marinelius}. There, it is observed that, by means of a general dimensional reduction technique, based on the classical 
gauge choices, spin-0 particles can be part of either massless or massive 
multiplets containing spin-1 in addition, i.e. of an either massless or massive  (1/2,1/2) multiplet, according to the gauge choice, thus suggesting a geometric interpretation of mass.

\section{The confining color--electric charge dipole potential}\label{sec3}

 An interesting perturbation of the free motion can be obtained from a solution to a certain Poisson equation on $S^3$, the one where the source is a color-electric charge dipole, as composed  by two opposite color-electric charge sources on $S^3$, placed in antipodal positions,
\begin{align}
& V_{CED}(\chi)= \Gamma_S(\chi) -\Gamma_N(\chi)
= -\alpha_sN_c\cot \chi +\lambda,\mbox{ with }\lambda=\lambda_S -\lambda_N\,,
\label{Wloops}\\
& \quad \Gamma_N(\chi)=\frac{\alpha_s N_c}{\pi}(\pi -\chi)\cot\chi +\lambda_N, \quad \Gamma_S=-
\frac{\alpha_sN_c}{\pi}\chi\cot\chi +\lambda_S,
\label{curved_dipole}
\end{align}
where the $\lambda$'s  are constants.
Here $\chi=\stackrel{\frown}{r}/R$, with $\stackrel{\frown}{r}$ the length of an arc of a great circle on $S^3$ in the $\chi$ variable, and $R$ is the hyper-spherical radius (interpreted as the compactification radius \cite{KTV}).  
In \cite{KC2016} this potential has been interpreted so that $\Gamma_N(\chi)$ and $\Gamma_S(\chi)$ represent the color-electric potentials due to single quarks of `negative' (anti-quark)  and `positive' (quark) color electric charges, placed at antipodal points of the hyper-sphere \cite{Brigita}. 
For the purposes of strong interaction physics  the  magnitude of the potential in (\ref{curved_dipole}) has been taken as the product of the strong coupling $\alpha_s$ with the number of colors $N_c$, as it  has been deduced in \cite{KC2016} from Wilson loops with cusps on $S^3$ following the method presented in \cite{Belitsky}.
 
Including the potential (\ref{Wloops}) into the free motion wave equation (\ref{free_on_S3}), we get
\begin{equation}
\left[-\Delta_{S^3}(\chi)  +1 -\alpha_sN_c\cot\chi  \right]
\psi_{K\ell }(\chi)=\left( (K+1)^2 -\lambda \right)\psi_{K\ell} (\chi),
\label{dipole_on_S3}
\end{equation}
A further transformation of the wavefunctions as in (\ref{1st_chng}),
\begin{equation}
\psi_{K\ell}(\chi)=\frac{u_{K\ell}(\chi)}{\sin \chi }\,,
\label{2nd_chng}
\end{equation}
leads to a Schr\"odinger-type equation,
\begin{eqnarray}
\left(-\frac{\partial ^2}{\partial \chi^2} + V_F(\chi)\right)u_{K\ell}(\chi)&=&(K+1)^2 u_{K\ell }(\chi).
\label{Schr_fndmntls}
\end{eqnarray}
Here,  $V_F(\chi)$ is given by
\begin{eqnarray}
V_F(\chi)&=&\frac{\ell(\ell +1) }{\sin^2\chi} -\alpha_sN_c\cot\chi +\lambda,
\label{tRM_undercover}
\end{eqnarray}
and will be referred to in the following  as the `one-dimensional color-electric charge dipole' potential. Notice that the sign in front of the cotangent function is inessential and can be reversed by interchanging $\Gamma_S(\chi)$ and $\Gamma_N(\chi)$ in (\ref{Wloops}).

\section{Potentials induced  by conformal deformations}\label{sec4}
As mentioned in the introduction, there is another procedure to obtain potentials of interest. Under a deformation of the metric on the $S^1\times S^3$ manifold by a conformal factor, here chosen as $e^{\mp 2f}$,
\begin{equation}
g_{S^1\times S^3}^*=e^{\mp 2f}g_{S^1\times S^3} \,,
\label{mtrc_dfrm}
\end{equation}
the conformal d'Alembertian $\Box_{g_{S^1\times S^3}}$ transforms by a similarity followed by a re-scaling
\begin{equation}
\Box_{g^\ast_{S^1\times S^3}}=e^{\pm 3f} \circ  \Box_{g_{S^1\times S^3}}  \circ e^{\mp f}\,=
e^{\pm 2f} \, e^{\pm f} \circ  \Box_{g_{S^1\times S^3}}  \circ e^{\mp f}\,.
\label{rsclng}
\end{equation}
We now analyze consequences of this fact.

\subsection{Induced scalar + gradient potentials in the quasi-radial variable }
We will consider scaling factors depending only on the quasi-radial variable $\chi$, while leaving intact the round metric in the 3D angular variables $\theta$ and $\varphi$. For this reason, we use (\ref{fctrz_wf}) and (\ref{Gegenb}) to pass from $\Box_{g_{S^1\times S^3}} (\tau, \chi,\theta,\varphi)$  in (\ref{Canz1}) to the reduced $\chi-$d'Alembertian,
\begin{eqnarray}
\Box_{g_{S^1\times S^3}}(\chi)&=&
 -(K+1)^2 -\frac{\partial^2}{\partial \chi^2} - 2\cot\chi \frac{\partial}{\partial \chi }+\frac{\ell (\ell +1)}{\sin^2\chi}+1 \ .
\label{Box_reduced}
\end{eqnarray}
In this way, the corresponding full  conformal wave operator  is converted to a reduced one given by,
\begin{eqnarray}
 \Box_{g_{S^1\times S^3}}(\chi)&=& -(K+1)^2 -\Delta_{S^3} (\chi) +1,
\label{Box_eq_def_rdc}
\end{eqnarray}
with $\Delta_{S^3}(\chi)$ already defined in (\ref{rdcd_dltS3}).
Then, the corresponding reduced conformal wave equation describing 
the coupling of a scalar massless particle to the  deformed metric becomes,
\begin{equation}
\Box_{g^\ast_{S^1\times S^3}}(\chi) \Phi _{K\ell}(\chi)=
e^{\pm 2f}\, e^{\pm f} \Box_{g_{S^1\times S^3}}(\chi)\, \left[e^{\mp f}\Phi_{K\ell}(\chi)\right]=0.
\label{reduced_dAl}
\end{equation}
Next we work out the first and second derivatives of $e^{\mp f}\Phi_{K\ell}(\chi)$ as
\begin{eqnarray}
\frac{\partial}{\partial \chi}e^{\mp f}\Phi_{K\ell}&=&
e^{\mp f}\left[\frac{\partial}{\partial \chi} \mp \frac{\partial f}{\partial \chi} \right]\Phi_{K\ell} ,\nonumber\\
\frac{\partial ^2}{\partial \chi^2}e^{\mp f}\Phi_{K\ell}&=&
e^{\mp f}\left[ \frac{\partial^2\Phi_{K\ell}}{\partial \chi^2}   
\mp 2\frac{\partial f}{\partial \chi}\frac{\partial \Phi_{K\ell}}{\partial \chi}
\mp\frac{\partial ^2 f}{\partial \chi^2}\Phi_{K\ell}
+\left( \frac{\partial f}{\partial \chi}\right)^2\Phi_{K\ell} \right],
\label{drv_fls}
\end{eqnarray}
where we temporarily suppress the functions arguments for the  ease of  notations. 
Now, with the aid of the above formulas and  taking the 
upper signs for concreteness, the $e^{-f}$ factor in (\ref{dragged})-(\ref{Box_eq_def_rdc}) can be dragged through $\Delta_{S^3}(\chi)$ from the very right to the very left, and be canceled out by $e^{f}$. In so doing,
(\ref{reduced_dAl}) simplifies to
\begin{eqnarray}
{e^{2f} \Big[} &-&(K+1)^2 -\Delta_{S^3}(\chi)+\mbox{Scal}_{g^\ast_{S^1\times S^3}}
+2\frac{\partial f}{\partial \chi}\frac{\partial }{\partial \chi}
{\Big]}\Phi_{K\ell}(\chi)=0,
\label{dragged}
\end{eqnarray}
where Scal$_{S^1\times S^3}^\ast$ denotes the re-scaled  curvature, calculated as
\begin{eqnarray}
\mbox{Scal}_{S^1\times S^3}^\ast &=& 2\frac{\partial f}{\partial \chi}\cot\chi - \left(\frac{\partial f}{\partial \chi} \right)^2  +\frac{\partial^2 f }{\partial \chi^2} +1.
\label{Scal_gast}
\end{eqnarray}  
Putting all together, the conformal wave equation in the re-scaled metric becomes
\begin{eqnarray}
e^{2f}\left[(K+1)^2 -\frac{\partial ^2}{\partial \chi^2} -2\cot\chi \frac{\partial }{\partial \chi} +\frac{\ell (\ell +1)}{\sin^2\chi}+{\hat V}_{CMF}(\chi) \right]
\Phi_{K\ell}(\chi)=0.
\label{indcd_ptl}
\end{eqnarray}
Here, ${\hat V}_{CMF}(\chi)$ is an operator potential of scalar + gradient type, induced by the metric deformation, which reads
\begin{eqnarray}
{\hat V}_{CMF}(\chi)&=&\mbox{Scal}_{g^\ast_{S^1\times S^3}}+2\frac{\partial f}{\partial \chi}\frac{\partial}{\partial \chi}\nonumber\\
&=& 2\frac{\partial f}{\partial \chi}\cot\chi - \left(\frac{\partial f}{\partial \chi} \right)^2  +\frac{\partial^2 f }{\partial \chi^2} +1+ 2\frac{\partial f}{\partial \chi}\frac{\partial}{\partial \chi}.
\label{grd_pt}
\end{eqnarray}
In this way, free stationary motion in the deformed metric, becomes  equivalent to stationary motion in the original one in the presence of the 
scalar+ gradient  potential (\ref{grd_pt}). Therefore, we have a way to induce potentials through conformal metric deformations. When using these potentials to model the internal dynamics of gauge boson exchange, they can be viewed as `mean field' potentials,  hence the acronym of `conformal mean field' in the notation ${\hat V}_{CMF}(\chi)$.

However, these potentials fail to  show up at the level of the Schr\"odinger equation because they vanish upon the change 
\begin{equation}
\Phi_{K\ell} (\chi)=\frac{U_{K\ell}(\chi)}{h(\chi)},
\label{nllfng_grd_trm}
\end{equation}
performed with the aim to remove the second derivative, 
$\left(-2\cot\chi +2 f^\prime\right)\frac{\partial }{\partial \chi} $. 
In  this case, a straightforward calculation shows that the condition for removing the first derivatives of the wave functions takes the form of
\begin{equation}
\frac{h'}{h}=\left( \ln h\right)'=\cot\chi -f'=\left( \ln \sin \chi -f\right)',
\label{drvtv_rmvl}
\end{equation}
whose solution  $h$ is given by
\begin{equation}
h=e^{-f}\sin\chi.
\label{g_fixed}
\end{equation}
With that choice, not only the gradient terms but the entire potential is removed, thus returning Schr\"odinger's version of  (\ref{dragged})  to (\ref{degeneracy}) and $\Phi_{K\ell}(\chi)$  to $U(\chi) =\sin\chi S_{K\ell}(\chi)$,
 the solution of the undeformed conformal wave equation in its  reduction to Schr\"odinger's equation. For this reason, we continue our explorations at the curved manifold.

\subsubsection{Exact solutions}

We are interested in the relation between the solutions to the deformed  d'Alembertian  $\Box_{g^\ast_{S^1\times S^3}}$, and those of the original one, $\Box_{g_{S^1\times S^3}}$. It is given by the following fact \cite{BirrelDavis},
\begin{equation}
  \Box_{(g^\ast=e^{\mp 2f}g)} u_{g^\ast}=\Box_{(g^\ast=e^{\mp 2f}g)} \left(
  e^{\pm f}u_{g}\right) =0 \quad \mbox{if\,\, and\,\, only\,\, if}\quad
 \Box_g\ (u_{g}) =0.
\label{fund_thrm}
\end{equation}
 This statement follows from the observation that upon  
substitution of (\ref{rsclng}) into the first equation in (\ref{fund_thrm}), the conformal wave equation on the re-scaled metric becomes equivalent to the d'Alembert equation on the orginal one, subjected to a similarity transformation by $e^{\pm f}$,  the inverse square root of the metric deformation factor in (\ref{mtrc_dfrm}), i.e. to
\begin{eqnarray}
\left[ e^{\pm f} \circ  \Box_{g_{S^1\times S^3}}  \circ e^{\mp f}\right] u_{g^\ast_{S^1\times S^3}}&=&0.
\label{sim_trnfrmde}
\end{eqnarray}
Similarly to (\ref{smlrt_trnsfm}), the latter equation implies
\begin{equation}
u_{g^\ast_{S^1\times S^3}}=e^{\pm f}u_{g_{S^1\times S^3}}.
\label{bong}
\end{equation}
For the specific case of a metric deformation in the quasi-radial variable alone, and upon making use of (\ref{fund_thrm}), the solutions $u_{g^\ast_{S^1\times S^3}}$, denoted by $\Phi_{K\ell}(\chi)$ in the preceding section, are therefore expressed as
\begin{equation}
\Phi_{K\ell}(\chi)=e^f S_{K\ell}(\chi), 
\label{fctr_stz}
\end{equation}
with $S_{K\ell}(\chi)$ corresponding to $u_{g_{S^1\times S^3}}$,  given in (\ref{Gegenb}). 
By the aid of (\ref{fctrz_wf})-(\ref{1d_Schr_ty}) we obtain the conformal wave equation on the deformed $S^1\times S^3$ metric in terms of a similarity transformation of the Casimir invariant of the $so(4)$ Lie algebra, 
\begin{equation}
\left[e^{\pm f} {\mathcal K}^2(\chi)e^{\mp f }\right] \left[e^{\pm f}S_{K\ell }(\chi)\right]= \left[ (K+1)^2 -1\right] \ \left[e^{\pm f}S_{K\ell }(\chi)\right]\,.
\end{equation}
Therefore, the solutions to the similarity transformed   Casimir invariant of the $so(4)$ Lie algebra are necessarily given by correspondingly re-scaled ultra-spherical harmonics (the quasi-radial parts of them in our case).
They are the new carrier spaces  of  the $so(4)$ Lie algebra in a new non-unitary equivalent representations. 
   In consequence,  all scalar+ gradient potentials induced by metric deformations turn out to be  exactly solvable. When only scalar potentials are considered, this is no longer true, but still a subclass of interesting solvable ones appear, see Subsection \ref{sec4.3}.

\subsection{Master Formula for the scalar pieces of induced potentials}
Equipped with the knowledge on the shape of the  solutions in (\ref{fctr_stz}), we now  elaborate a bit more on (\ref{indcd_ptl}), (\ref{grd_pt}) by rewriting the last term in (\ref{grd_pt}) according to
\begin{equation}
2\frac{\partial f}{\partial \chi}\frac{\partial}{\partial \chi}\left[e^f S_{K\ell}(\chi)\right]=
e^f\left[ 2\frac{\partial f}{\partial \chi}\frac{\partial}{\partial \chi} 
 +2\left(\frac{\partial f}{\partial \chi}\right)^2\right] S_{K\ell}(\chi).
\label{DK_oprtr}
\end{equation}
Back substitution of the latter equation  into  (\ref{indcd_ptl}), and upon addition and subtraction ( for the sake of later convenience) of the term,
$2\frac{\partial f}{\partial \chi}e^fK\cot \chi $, amounts  to
\begin{eqnarray}
\left[ -(K+1)^2 -\frac{\partial ^2}{\partial \chi^2}-2\cot\chi \frac{\partial }{\partial \chi} +
\frac{\ell(\ell+1)}{\sin^2\chi}+1 +\left(\frac{\partial f}{\partial \chi}\right)^2 +\frac{\partial ^2f}{\partial \chi^2}+2\frac{\partial f}{\partial \chi}(K+1)\cot\chi 
\right]&&\nonumber\\
\times  \left[e^f S_{K\ell^\prime}(\chi)\right]
+e^f \, 2\frac{\partial f}{\partial \chi } \left( \frac{\partial} {\partial \chi} -K\cot\chi\right) S_{K\ell^\prime}(\chi)=0.&&\nonumber\\
\label{path_1}
\end{eqnarray}
Notice that in the process,  the sign in front of $\left(\frac{\partial f}{\partial \chi}\right)^2$ 
got changed from negative in (\ref{grd_pt}) to positive in the latter equation  due to its partial cancellation with the second term in (\ref{DK_oprtr}). \\
\noindent
Now it can be verified through a direct calculation that for the maximal allowed value of the angular momentum,  $\ell_{max} =K$, the  gradient term re-designed in the above way, has the following property,
\begin{equation}
\left(\frac{\partial}{\partial \chi}-K\cot\chi \right)S_{K K}(\chi)=0.
\label{lwst_hrrch}
\end{equation}
Substitution of (\ref{lwst_hrrch}) into (\ref{path_1}) shows that
on the space of the $e^fS_{KK}(\chi)$  functions,  this gradient term drops out from (\ref{path_1}) leaving us with 
 \begin{eqnarray}
\left[-(K+1)^2 -\frac{\partial ^2}{\partial \chi^2}-2\cot\chi \frac{\partial }{\partial \chi} +\frac{K(K+1)}{\sin^2\chi}+1 +\left(\frac{\partial f}{\partial \chi}\right)^2 +\frac{\partial ^2f}{\partial \chi^2}+2\frac{\partial f}{\partial \chi}(K+1)\cot\chi 
\right]&&\nonumber\\
\times  \left[e^f S_{KK} (\chi)\right]=0,&&\nonumber\\
\label{path_2}
\end{eqnarray}
an equation whose ground state wave function is that very same, $e^fS_{KK}(\chi)$, equivalently, $\Phi_{KK}(\chi)$ from (\ref{fctr_stz}).
Eliminating the first derivative from the latter equation through the replacement,
\begin{equation}
\Phi_{K K }(\chi) =\frac{U_{KK}(\chi)}{\sin \chi}, \quad \mbox{i.e.} \quad U_{KK}(\chi)=
e^f\sin\chi S_{K K}(\chi)=e^f\sin^{K+1}\chi,
\label{U-sltn}
\end{equation}
is now standard and leads to,
\begin{eqnarray}
\left[-\frac{\partial ^2}{\partial \chi^2} +V^{(K)}_{I}(\chi)
\right]  U_{KK}(\chi)&=&(K+1)^2U_{KK}(\chi).
\label{path_3}
\end{eqnarray}
Here, $V_{I}^{(K)}(\chi)$ denotes a one dimensional potential induced by metric
 deformations according to (\ref{mtrc_dfrm}), now a scalar function and no longer an operator, and for which the function $f$ depends on the quasi-radial variable $\chi$ alone. It reads,
\begin{eqnarray}
V_{I}^{(K)}(\chi) &=&\frac{K(K+1)}{\sin^2\chi} +\left(\frac{\partial f}{\partial \chi}\right)^2 +\frac{\partial ^2f}{\partial \chi^2}+2\frac{\partial f}{\partial \chi}(K+1)\cot\chi\, ,
\label{Master_Formula_Pre}
\end{eqnarray}
see comment after (\ref{path_1}).\\

\noindent
It is our goal now to study as to what extent such potentials are meaningful on their own rights for all allowed values of the angular momentum, i.e. upon the
replacements $K(K+1)/\sin^2\chi\to \ell(\ell+1)/\sin^2\chi$, and also possibly $(K+1)\cot\chi\to (\ell +1)\cot\chi$, for $\ell=0,1,2..., K$. Stated differently, we ask the question whether extensions of the space of functions in (\ref{path_3}) are possible, in which the scalar pieces of the potentials in 
(\ref{path_1}), generalized in the way just explained,  remain exactly solvable. For this purpose we introduce  the following Hamiltonians,
 \begin{eqnarray}
H^{(\ell)}_I U_{K\ell}(\chi)\equiv \left[-\frac{\partial ^2}{\partial \chi^2} +V^{(\ell )}_{I}(\chi) -(K+1)^2
\right]  U_{K\ell }(\chi)&=&0,
\label{pre_1}
\end{eqnarray}
with the potential being given by the following Master Formula
\begin{eqnarray}
V_{I}^{(\ell )}(\chi) &=&\frac{\ell (\ell +1)}{\sin^2\chi} +\left(\frac{\partial f}{\partial \chi}\right)^2 +\frac{\partial ^2f}{\partial \chi^2}+2\frac{\partial f}{\partial \chi}(K+1)\cot\chi,\nonumber\\
&=&\frac{\ell (\ell +1)}{\sin^2\chi }+\mbox{Scal}^\ast_{S^1\times S^3} +2\left( \frac{\partial f}{\partial \chi}\right)^2.
\label{new_class}
\end{eqnarray}
The $V_{I}^{(\ell )}(\chi)$ will be called 
`one-dimensional induced scalar potentials' .  Essentially they describe the coupling of a massless scalar particle to the scalar curvature on the deformed metric.
Furthermore, the wave functions $U_{K\ell}(\chi)$ will be written in the form  
\begin{equation}
U_{K \ell}(\chi) =U_{KK}(\chi)t_n^{\alpha_K,\beta_K}(\chi)=e^{f}\sin^{K+1}\chi \, t^{\alpha_K\beta_K}_n(\chi),\quad n=K-\ell,
\label{Q_3}
\end{equation}
with $t_n^{\alpha_k,\beta_K}(\chi)$ being some unknown functions, to be found upon solving (\ref{new_class}).

\subsection{Examples of one-dimensional induced scalar potentials}\label{sec4.3}
In this section we illustrate the ideas developed above by analyzing some examples of potentials, being three of them exactly solvable, and another one only quasi-exactly solvable.

\subsubsection{The trigonometric Rosen-Morse potential}
Let us apply the procedure presented in the preceding section to the case of a conformal factor whose argument is linear in the quasi-radial variable,
\begin{eqnarray}
g^\ast_{S^1\times S^3}&=&e^{-\alpha_K\chi} g_{s^1\times S^3}, \quad \alpha_K=\frac{\alpha_s N_c}{K+1},
\label{our_gstar}
\end{eqnarray}
i.e. 
\begin{equation}
f=\frac{\alpha_K\chi}{2}, \quad \frac{\partial f}{\partial \chi} =\frac{\alpha_K}{2}, \quad \frac{\partial ^2f}{\partial \chi^2}=0.
\label{f_fnctn}
\end{equation}
In this case, one encounters the following induced equation,
\begin{eqnarray}
H_I^{(\ell)}(\chi) U_{K\ell}(\chi)&\equiv& \left[-\frac{\partial ^2}{\partial \chi^2} +V^{(\ell )}_{I}(\chi) -(K+1)^2
\right]  U_{K\ell }(\chi)=0,\nonumber\\
V_I^{(\ell)}(\chi)&=& \frac{\ell (\ell+1)}{\sin^2\chi} +\alpha_K (K+1)\cot\chi +\frac{\alpha_K^2}{4}.
\label{tRM_KK}
\end{eqnarray}
 
The wave functions in (\ref{Q_3}) are then given by
\begin{equation}
U_{K \ell}(\chi) =U_{KK}(\chi)t_n^{\alpha_K,\beta_K}(\chi)=e^{\frac{\alpha_K\chi}{2}}\sin^{K+1}\chi \, t^{\alpha_K\beta_K}_n(\chi).
\label{tRM_1_1}
\end{equation}
Comparison of (\ref{tRM_KK}) to (\ref{tRM_undercover}) shows that for 
\begin{equation}
\alpha_K(K+1)=-\alpha_sN_c, \quad \lambda=\frac{\alpha_K^2}{4}, \quad \ell=0,1,2,..., K,
\label{speci_lambda}
\end{equation}
the potential induced by the simplest metric deformation in (\ref{our_gstar}) equals the one obtained in Section \ref{sec3} (the color--electric charge dipole),
\begin{equation}
V_{F}(\chi)=V_{I}^{(\ell)}(\chi).
\end{equation}
Moreover, these equal potentials  represent (modulo the additive constant) a particular case of the trigonometric Rosen-Morse potential, denoted by $V_{tRM}(\chi)$ and given by \cite{Khare},
\begin{equation}
V_{tRM}(\chi)=\frac{a(a -1)}{\sin^2\chi} +2b\cot\chi,
\label{tRM_stndrd}
\end{equation}
namely the one in which the $a$ parameter has been replaced  by  $a=\ell + 1$ with $\ell$ non-negative integer. 
By this observation, the well studied trigonometric Rosen-Morse potential could be re-interpreted in a two-fold way, firstly as the simplest representative of the potentials of the new family in (\ref{new_class}), and secondly as the 
solution to a Poisson equation on $S^3$, as in (\ref{tRM_undercover}), in combination with (\ref{speci_lambda}). Recall that the latter was the color-electric charge dipole potential. Though similar in form, these two interactions are distinguished by the nature of their sources. The induced potentials mimic the overall gauge field dynamics of the systems, while the fundamental one is attributed to a charge dipole with the charges being placed at antipodal positions. Correspondingly, they are apt for the description of distinct systems, two-body in  the first case, versus $2n$ body with $n>1$ in the second. This distinction will acquire relevance in Section 5 below, where the second interpretation of the trigonometric Rosen-Morse potential will be preferred.

Now, the general procedure to check whether or not the potential is exactly solvable  is to analyze the solubility of the equation satisfied by the $t^{\alpha_K\beta_K}_n(\chi)$ function.
For that purpose one proceeds as follows (see \cite{ShiDong} and references therein). Introduce an appropriate variable  change, in our case $z=\cot\chi$, and rewrite (\ref{tRM_1_1}) in the new variables as
\begin{eqnarray}
 U_{K \ell}(\chi)\to {\mathcal U}_{K\ell}(z)&=&\phi (z) u(z),\label{vrchng_1}\\
\phi(z)&=&e^{\frac{\alpha_K\cot^{-1}z}{2}}\frac{1}{(1+z^2)^{\frac{K+1}{2}}}.
\label{phi_z}
\end{eqnarray}
Next cast the potential in (\ref{tRM_KK}) in the new variable to get,
\begin{eqnarray}
V_I^{(\ell)}(\chi)\to {\mathcal V}(z)&=& \ell (\ell+1)(1+z^2) +\alpha_K (K+1)z  +\frac{\alpha_K^2}{4}.
\label{tRM_KK_z}
\end{eqnarray}
Now the corresponding one-dimensional Schr\"odinger equation  
\begin{equation}
\left[-\frac{1}{(1+z^2)^2}\frac{{\mathrm d}^2}{{\mathrm d}^2z} +{\mathcal V}(z) \right]
{\mathcal U}_{K\ell}(z)=E\,  {\mathcal U}_{K\ell}(z), \quad E=(K+1)^2,
\label{Q_2}
\end{equation}
can be reduced to a second order differential equation of the type
\begin{equation}
u''  +g u' +h u=0,
\label{Q_4}
\end{equation}
where  the functions $g$ and $h $ are defined in \cite{ShiDong} as 
\begin{eqnarray}
g&=&2\frac{\phi '}{\phi } +\frac{\rho'}{\rho}, \nonumber\\
h &=&\frac{\phi ''}{\phi } +\frac{\phi '}{\phi }\frac{\rho'}{\rho} +\frac{E-{\mathcal V}}{\rho^2}.
\label{Q_5}
\end{eqnarray}
Here, $\rho=\frac{{\mathrm d}\cot\chi }{{\mathrm d}\chi}=-1/\sin^2\chi=-(1+z^2)$, 
while ${\mathcal V}(\chi)$ is the potential in (\ref{Q_22}) expressed  by means of the new variable $z$ (temporarily suppressed in the latter equation).
Next we calculate the expressions in (\ref{Q_5}), beginning with $g$.
For that purpose we first calculate
\begin{equation}
\frac{\phi'}{\phi}=\frac{1}{1+z^2}\left[-\frac{\alpha_K}{2} -(K+1)z\right],
\label{Q_7}
\end{equation}
and
\begin{equation}
\frac{\rho'}{\rho}=\frac{2z}{1+z^2},
\label{Q_8}
\end{equation}
Substitution of the last two expressions into the first equation in
(\ref{Q_5}) leads to
\begin{equation}
g=\frac{(-1)}{1+z^2}\left[ \alpha_K  + 2K z\right] 
\label{Qgz}
\end{equation}
As a preparation to the calculation of $h$ we first work out
\begin{eqnarray}
\frac{E-{\mathcal V}}{\rho^2}=\frac{1}{(1+z^2)^2}
\left[(K+1)^2 -\ell (\ell +1)(1+z^2)  -\alpha_K (K+1)z  -\frac{\alpha_K^2}{4}
\right]&&\label{E_V}
\end{eqnarray}
Finally, we calculate $\frac{\phi ''}{\phi }$ as
\begin{eqnarray}
\frac{\phi''}{\phi}=
\left(\frac{\alpha_K}{2} +(K+1) z\right)
\frac{1}{(1+z^2)^2}\left[ \frac{\alpha_K}{2}+(K+3)z \right] -\frac{(K+1)}{1+z^2}\,.
\label{QFdblprm_f}
\end{eqnarray}
Substitution of equations  (\ref{Q_7}), (\ref{Q_8}), (\ref{E_V}), and  (\ref{QFdblprm_f}) into the second
equation in (\ref{Q_5}) yields
\begin{equation}
h=\frac{K(K+1)-\ell(\ell +1)}{1+z^2}.
\label{Q_hz}
\end{equation}
With that, the second order equation in (\ref{Q_4}) adopts the form
\begin{equation}
(1+z^2) u'' + (1+z^2)g u' +(K(K+1)-\ell (\ell +1))u=0,
\label{Q_u}
\end{equation} 
with
\begin{equation}
(1+z^2)g(z)= -\alpha_K -2Kz.
\label{def_T}
\end{equation}
The equation in (\ref{Q_u}) coincides  with Romanovski's hyper-geometric 
equation, with  $(1+z^2)g(z)$ being equal to a Romanovski polynomial of first degree according to 
\begin{equation}
(1+z^2)g(z)=-\alpha_K\cot^{-1}z -2 K z= R^{-\alpha_K,\beta_K}_1(z).
\label{Rdrs_4}
\end{equation}
The Romanovski  polynomials  are obtained by means of the Rodrigues formula from the  following weight function,
\begin{equation}
  \omega^{\alpha,\beta}(x)=(1+x^2)^{\beta-1} \exp(-\alpha \cot^{-1}x),
   \end{equation}
(see  \cite{Raposo} for a review). Their parameters for the case of our interest, that in which they constitute an infinite set of orthogonal polynomials, have been determined in \cite{TQC} as 
\begin{equation}
\alpha_K=\frac{\alpha_sN_c}{K+1}, \quad \beta_K=-K, \quad K=n+\ell.
\label{plnmls_cnsts}
\end{equation}
Therefore, the simplest induced potential turns out to be  exactly solvable and equal to the trigonometric Rosen-Morse potential.

\subsubsection{The trigonometric P\"oschl-Teller potential}
A further potential is obtained from the choice
\begin{eqnarray}
f(\chi)=\frac{\alpha}{2}\ln \cos\chi, &\quad& \frac{\partial f(\chi)}{\partial \chi}= -\frac{\alpha}{2}\tan\chi, \nonumber\\
\left( \frac{\partial f(\chi)}{\partial \chi} \right)^2=
\frac{\alpha^2}{4}\left( \frac{1}{\cos^2\chi} -1\right), &\quad&
 \frac{\partial^2f(\chi)}{\partial \chi^2}=-\frac{\alpha}{2}\frac{1}{\cos^2\chi}.
\label{f_PTI}
\end{eqnarray}
Substitution into the Master Formula (\ref{new_class}) yields
\begin{eqnarray}
V_{PTI}(\chi):=V_I^{(\ell)}-(\ell +1)^2&=&\frac{\ell(\ell +1) }{\sin^2\chi} +\frac{\frac{\alpha}{2} \left(\frac{\alpha}{2} -1\right)}{\cos^2\chi} -\left( \ell +1+\frac{\alpha}{2}\right)^2.
\label{PTI_1}
\end{eqnarray}
This expression reproduces  precisely the exactly solvable  trigonometric P\"oschl-Teller potential, $V_{PTI}(\chi)$ (P\"oschl-Teller I)  as presented in SUSY-QM \cite{Khare}, with the additive constant term setting the energy of the ground state to equal zero.

\subsubsection{The trigonometric Scarf potential}
As another example we consider the function
\begin{eqnarray}
f(\chi)=\frac{\alpha}{2}\ln |\csc\chi +\cot\chi|, &\quad& \frac{\partial f(\chi)}{\partial \chi}= \frac{\alpha}{2}\frac{1}{\sin\chi}, \nonumber\\
\left( \frac{\partial f(\chi)}{\partial \chi} \right)^2=
\frac{\alpha^2}{4}\csc^2\chi , &\quad&
 \frac{\partial^2f(\chi)}{\partial \chi^2}=-\frac{\alpha}{2}\csc\chi \cot\chi.
\label{f_ScI}
\end{eqnarray}
Substitution into the Master Formula (\ref{new_class}) yields
\begin{eqnarray}
V_{ScI}(\chi):=V_I^{(\ell)}-(K+1)^2&=&{\Big[}\ell (\ell +1)
 +\frac{\alpha^2}{4}{\Big]}\csc^2\chi\nonumber\\
& +& \frac{\alpha}{2}(2\ell+1)\csc\chi \cot\chi -(K +1)^2.
\label{Scrf_1}
\end{eqnarray}
This expression reproduces precisely the exactly solvable  trigonometric Scarf  potential, $V_{ScI}(\chi)$, (Scarf I)  as considered in SUSY-QM \cite{Khare}, with the additive constant term setting the energy of its ground state to equal zero. In this way, all three exactly solvable trigonometric potentials known from SUSY-QM have been shown to be describable by conformal metric deformations of the compactified Minkowski spacetime. Notice, though, that while all three of them have ground state wave functions as  parts of $so(4)$  carrier spaces, and therefore are all $so(4)-$symmetric, only in  the trigonometric Rosen-Morse potential, this symmetry extends to the excited states. 

\subsubsection{The potential of the MIC-Kepler problem\footnote{This result has been obtained after submission of galley proofs, so it does not appear in the published version.}
}

Combining the conformal metric deformation for the trigonometric Rosen-Morse with the one for the P\"oschl -Teller potential (though with $\sin\chi$ in place of $\cos\chi$ in the latter) i.e. by
$f(\chi)=\beta/2\ln \sin\chi +\alpha_K\chi/2$, we obtain 
\begin{equation}
V_{MIC-K}(\chi)=\frac{\ell(\ell +1)}{\sin^2\chi } +\frac{\mu^2}{\sin^2\chi}+\alpha_K(K+1)\cot\chi -\frac{\beta^2}{4}
 +\frac{\alpha_K^2}{4}-\beta(K+1) -(K+1)^2,
\end{equation}
with $\beta $ constant and $\mu^2=\beta/2(\beta/2 -1)+ \beta(K+1)$.
This is the  exactly solvable MIC-Kepler potential describing on $S^3$, according to \cite{Gritsev}, the  interaction of  Schwinger's dyons \cite{Schwinger}, particles equipped by both electric  and magnetic monopole charges, in which case $\mu$ is integer or semi-integer. The present consensus is that quarks can not be dyons because the electric charge of a dyon has to be integer, versus  semi-integer in quarks. Our model shows that there exists a metric on the compactified Minkowski spacetime  which can give rise to a mean field potential of the dyon--dyon type. As to what extent such a metric could be relevant to physics, this can be decided only upon comparison of its predicted hadron magnetic form factors to data. 

\subsubsection{A quasi-exactly solvable potential}
In the case of $f$ having a quadratic dependence on the quasi-radial variable,
\begin{eqnarray}
g^\ast_{S^1\times S^3}&=&e^{-\alpha_K\chi^2} g_{s^1\times S^3},,
\label{b_2__gstar}
\end{eqnarray}
i.e. 
\begin{equation}
f=\frac{\alpha_K}{2}\chi^2, \quad \frac{\partial f}{\partial \chi}=\alpha_K\chi,
\quad \frac{\partial ^2f }{\partial \chi^2}=\alpha_K,
\label{f_fnctn_2}
\end{equation}
the resulting potential is
\begin{equation}
V_I^{(\ell)}(\chi)=\frac{\ell(\ell +1)}{\sin^2\chi} +2\alpha_K (K+1)\chi \cot\chi +\alpha_K^2\chi^2 +\alpha_K.
\label{Q_1}
\end{equation}
Upon  the change of $\chi$ to the new variable $z=\cot\chi$, the potential in (\ref{Q_1}) becomes,
\begin{equation}
{\mathcal V}(z)=\ell (\ell +1) (1+z^2) + \alpha^2_K(\cot^{-1}z)^2 +
\alpha_K +2\alpha_K(K+1)z\cot^{-1}z. 
\label{Q_22}
\end{equation}
Next we express the related function $U_{K\ell}(\chi)$ from (\ref{Q_3}) in terms of $z$
as
\begin{eqnarray}
U_{K\ell}(\chi)\to {\mathcal U}_{K\ell}(z)&=&\phi(z)u^{\alpha_K,\beta_K}_n(z),\nonumber\label{Gl_9}\\
\phi (z)&=& e^{\frac{\alpha_K(\cot^{-1}z)^2}{2}}\frac{1}{(1+z^2)^{\frac{-\beta_K+1}{2}}}.
\label{Q_6}
\end{eqnarray}
Reducing along the lines of the preceding  subsection (4.3.1) the equation in (\ref{tRM_KK_z}) corresponding to (\ref{Q_1}), 
we find that the $u$ functions are supposed to satisfy  the following  equation, 
\begin{equation}
(1+z^2) u'' + (1+z^2)g u' +(K(K+1) -\ell(\ell +1))u=0,
\label{Q_u_u}
\end{equation} 
with
\begin{equation}
(1+z^2)g =-2\alpha_K\cot^{-1}z -2 K z=
-2\alpha_K \chi +2\beta_K z.
\label{Rdrs_4_4}
\end{equation}
Curiously, this equation coincides in shape with Romanovski's equation (\ref{Q_u})-(\ref{def_T}), however the place of the Romanovski polynomial of first degree finds itself  occupied by the function in (\ref{Rdrs_4_4}). This equation has only one solution, namely the one corresponding to $K=\ell=0$, in which case 
$u$ is a  constant. This is precisely the case which corresponds  to the ground state  $U_{KK}(\chi)$.
The latter observation is illustrative of the quasi-exact solubility of the 
potential under consideration. Thus, the family of potentials in (\ref{new_class}) contains both exactly and quasi-exactly solvable potentials.

\section{Application to quark deconfinement}\label{sec5}

In the preceding exposition, we have considered a fixed compactificacion raduis $R$. It is natural to
consider the  asymptotic behavior for increasing values of $R$.
We will focus our attention on the trigonometric Rosen-Morse potential, viewed 
as the one-dimensional color--electric charge dipole potential  as in (\ref{tRM_KK}) (equivalent to   (\ref{tRM_undercover}) for $\lambda=\alpha_K^2/4$), associated to a physical color--electric charge dipole source, and we will show that it is relevant not only to the confinement but also to the deconfinement of quarks.

\subsection{The conformal wave equation with the one dimensional color-electric charge dipole  potential in units of MeV$^2$ }

In order to switch to dimensional equations, one possibility consists in  multiplying 
$\alpha_sN_c \cot\chi$ in (\ref{tRM_KK})  by $\frac{\hbar c}{R}\Lambda_{QCD}$ (with $\Lambda_{QCD}$ the QCD scale parameter), with the aim of keeping the 
parameters  as close as possible to the fundamental ones,
and the rest of the terms by
$\hbar^2c^2/R^2$. In doing so, one arrives at
\begin{eqnarray}
\frac{\hbar^2c^2}{ R^2}\left[
-\frac{{\mathrm d}^2}{{\mathrm d}\chi^2} +\frac{\ell (\ell +1)}{\sin^2\chi}
 \right]U_{K\ell}(\chi) +\alpha_sN_c \, \frac{\hbar c}{R}\Lambda_{QCD} \cot\chi 
 U_{K\ell}(\chi)
&=&{\mathcal E}^2_{K} U_{K\ell}(\chi),
\label{Gl_6_2}
\end{eqnarray}  
where the squared energy is given by,
\begin{equation}
{\mathcal E}^2_{K}=\frac{\hbar^2c^2}{ R^2}(K+1)^2 -
\frac{\alpha_s^2N_c^2\Lambda_{QCD}^2}{4} \frac{1}{(K+1)^2}\,.\label{Enrg_tRM_2}
\end{equation}
Formally, equation (\ref{Gl_6_2}) is the same as the `quadratic'  Schr\"odinger equation  with a scalar trigonometric Rosen-Morse potential.
Therefore, this phenomenological potential can be deduced from general considerations within the context of conformal wave operators on $M^c$, thus emphasizing its fundamental nature.

\subsection{The role of the `reduced mass-parameter' in Schr\"odinger's equation}\label{sec52}

Next, we want to introduce natural units in our problem, namely, units of MeV. To this end,   we divide (\ref{Gl_6_2}) by $\Lambda_{QCD}$, so this magnitude can now be parameterized as, 
\begin{equation}
\Lambda_{QCD}=2\mu_q c^2\,,
\label{mss_prmtr}
\end{equation}
where $\mu_qc^2$ formally looks like as a `reduced mass', though without having such physical meaning. 
To be specific, the Schr\"odinger version of (\ref{Gl_6_2}) now reads
\begin{eqnarray}
\frac{\hbar^2c^2}{2\mu_q c^2 R^2}\left[
-\frac{{\mathrm d}^2}{{\mathrm d}\chi^2} +\frac{\ell (\ell +1)}{\sin^2\chi}
 \right]U_{K \ell}(\chi) -\frac{\hbar c}{R}\alpha_sN_c\cot\chi U_{K\ell}(\chi)
&=&E_{K} U_{K \ell}(\chi)\,,
\label{Gl_6}
\end{eqnarray}  
where  $E_{K}={\mathcal E}^2_K/\Lambda_{QCD}$  denotes the energy in units of MeV, given by
\begin{equation}
E_{K}=\frac{\hbar^2c^2}{2\mu_q c^2 R^2}(K+1)^2 -\frac{\mu_q c^2 \alpha_s^2 N_c^2}{2}\frac{1}{(K+1)^2}\,.\label{Enrg_tRM}
\end{equation}
The energies resulting from this equation are discrete, and their number is infinite due to the infinite depth of the potential. They can be equivalently written as \footnote{Energies of this kind  have first been calculated by Schr\"odinger \cite{Schr41}, though within the cosmological context of a  H Atom  on Einstein's closed Universe,  $R^1\times S^3$, in which case $\alpha_s N_c$ 
is replaced by $\alpha Z$, and instead of $\mu_qc^2$, the reduced mass of the proton-electron system is used. The parameterization of the energy chosen in 
(\ref{curved_spctrm}) parallels in the strong interaction sector Schr\"odinger's formula from the electromagnetic sector.},
\begin{align}
\begin{split}
E_{K}=Ry^s\left[-\frac{1}{(K+1)^2} +\left(\frac{\left(a_0^q\right)^\ast }{R}\right)^2(K+1)^2
\right]\,, \\
\mbox{where } Ry^s=\frac{\mu_q c^2 \alpha_s^2N_c^2}{2}=\frac{\hbar ^2 c^2}{2\mu_qc^2 (\left(a_0^s\right)^\ast )^2}\,.
\end{split}
\label{curved_spctrm}
\end{align} 
Notice that $Ry^s$ coincides in shape with  the Rydberg constant in the H Atom, with the fine structure constant $\alpha$ replaced by $\alpha_s$, and the charge number $Z$ replaced by $N_c$, the number of colors in QCD. We will refer to it as the `strong Rydberg constant'.
Also, the parameter $\left(a_0^q\right)^\ast $, defined as
\begin{equation}
\left(a^q_0\right)^\ast =\frac{\hbar c}{\mu_q c^2 \alpha_s N_c}=
\frac{\lambda^s_c\!\!\!\!\!\!/}{\alpha_s N_c}, \quad \lambda^s_c\!\!\!\!\!\!/=\frac{\hbar c}{\mu_qc^2},\quad \mu_qc^2=\Lambda_{QCD},
\label{Bohr_rds}
\end{equation}
is the formal quarkish analogue of Bohr's reduced radius of the electron. Furthermore, $\lambda_c^s\!\!\!\!\!\!/$ \,\,   denotes the reduced Compton wave length of a quark. Moreover, the excited hadron levels present the same $(K+1)^2$-fold degeneracy of $|K\ell m>$ states inside a level, 
as those of the H Atom, thus reflecting the (dynamical) conformal $SO(2,4)$ symmetry of strong interaction.

In summary, we can conclude that  equations (\ref{Gl_6}) and (\ref{Gl_6_2}) have their root in the conformal wave operator on $S^1\times S^3$,  following (\ref{fund_thrm}), and thus are not phenomenological choices.  The most important new insight gained in the process is that they still refer to motions of a massless scalar particle on $S^1\times S^3$, because the $\mu_q c^2$ parameter in 
(\ref{mss_prmtr}) is not a physical reduced mass but the artifact of the presence of the $\Lambda_{QCD}$ scale parameter in (\ref{Gl_6_2}). They become phenomenological when considering $\mu_q$ as a free parameter.

\subsection{Application  to hadron spectra}

\begin{figure}[h]
  \begin{center}
  \includegraphics[width=9.5cm]{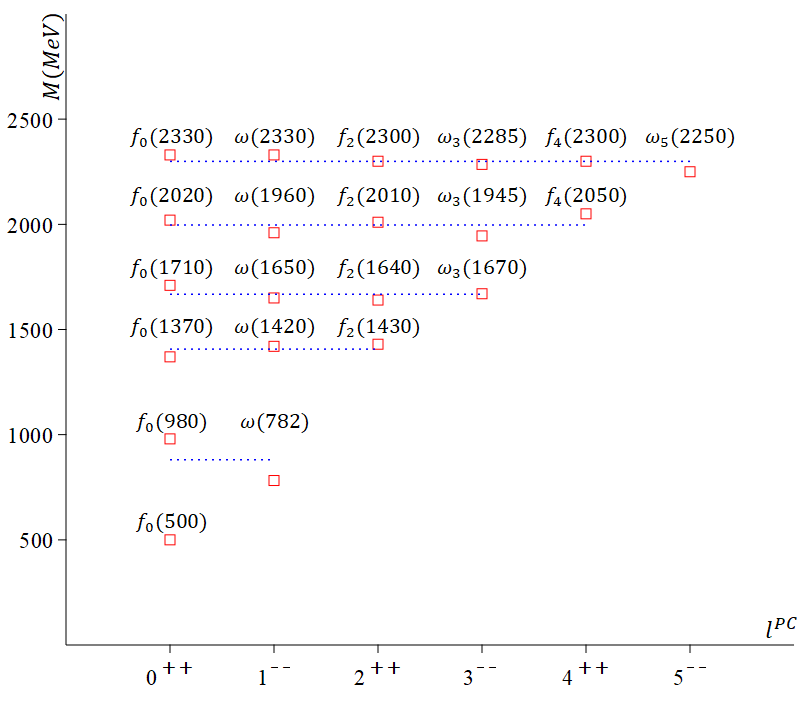}
  \caption{ The observed spectrum of the $f_0$ mesons}\label{fig01}
\end{center}
  \end{figure}

  Hadronic spectra, both baryonic and mesonic,  fit quite well into the above dynamical  scheme (see Fig.~\ref{fig01}  for a representative example, one among many) and thus support the conformal symmetry of strong interaction in the infrared, observed through the `freezing' of the strong coupling in this regime.

In that regard, it is now quite insightful to  compare eqs.~(\ref{Enrg_tRM_2}) and ~(\ref{Enrg_tRM}) to data.
 For example, the level splitting in the $f_0$ spectrum in Fig.~\ref{fig01}, among  other meson families,  have been adjusted in \cite{KC2016} by means of the formula (\ref{Enrg_tRM_2}) and the numerical values of the parameters have been found as
  \begin{equation}
\frac{\hbar ^2c^2}{R^2}=0.10964 \, \mbox{GeV}^2, \quad 
\frac{\alpha_s^2 N_c^2 \Lambda_{QCD}^2}{4}=1.0434 \,\, \mbox{GeV}^2.
\label{data_fits}
\end{equation}
The values of their positions have then be adjusted by adding to (\ref{Enrg_tRM_2}) the constant term, $C(R)=1.1873$ GeV$^2$.

Equations  (\ref{Gl_6}) and (\ref{Gl_6_2}) have been employed, prior to 
\cite{KTV} and the present work, as a phenomenological tool in several key data analyses on hadron spectroscopy ranging from meson and baryon spectra \cite{KC2016},\cite{Ahmed}  electromagnetic form factors \cite{MK_Formfactors} (treated within the framework of Dirac equation on $S^3$)  up to phase transitions \cite{AramDavid}, all of which have revealed the adequacy of the approach (also see \cite{LT14} for a concise summary). \footnote{Admittedly, in the quoted works the magnitude of the cotangent potential has not been taken as $\hbar c \alpha_sN_c/R$ as in (\ref{Gl_6}). Rather, it has been treated as a parameter, writing it as $2G/R$ with $G $ being given in units of MeV.fm, and fixed by data fit. Specifically in \cite{TQC} G=204.08 MeV.fm, $R$=2.31 fm, while the `reduced mass' parameter  takes the value of  $\mu c^2$=209.166 MeV.} Only in \cite{PRD10}  the potential has been included into the Klein-Gordon wave operator, however as a gauge potential. In doing so, the correct orderings of the $S_{n1/2}$ and $P_{n1/2}$ states in a level could be well reproduced as en effect of the kinematic $\Delta  \ell=1$ splitting.

The degeneracies in the hadron spectra have been independently addressed by several authors, more recent works being \cite{Afonin}, \cite{Afonin3}, \cite{Valcarse}. Specifically in \cite{Afonin}, \cite{Afonin3}, the classification of numerous data on light flavor mesons reported by the Crystal Barrel collaboration have been  analyzed for the first time within the context of conformal symmetry, though not in terms of levels but  in terms of Regge trajectories.
In \cite{Valcarse} a  screened confinement potential ansatz of the form 
$V_{sb}(r)=\left( \sigma r -\frac{{\bar \lambda}_q}{r}\right) \left(1 -e^{\frac{\nu}{r}} \right)$ has been employed in the meson data analyses and clustering in several mass regions observed. For example, in the mass-region around 2258 $\pm $38 MeV, the approximate degeneracy of the isovector $a_3(2275)$, $a_2(2255)$, $a_4(2255)$ mesons of positive PC parity, and the isovector $b_3(2245)$ meson of negative PC parity  has been reported, among many similar other groups of mass degenerate mesons. This type of degeneracies are well described too by the model here presented \cite{KC2016}, with the differences that {i) in the latter reference mesons with equal PC parties have been classified as degenerate and  placed in same level, 
(ii) somewhat bigger mass splitting among the states than in \cite{Valcarse} were allowed, though significantly smaller than the splittings between the levels.\\

\noindent
In the subsequent section we address the question on the deconfinement phenomenon with the method here advocated.

\subsection{The color electric charge dipole potential at asymptotically large compactification radii}
The energy formula in (\ref{curved_spctrm}) shows an interesting behavior in the large $R$ limit. Namely, assuming that the radius depends linearly on the temperature, the  limit  $\lim _{T\to \infty }R(T)\to \infty$ of the equation 
 (\ref{curved_spctrm}) can be considered,  finding that for an approximate constant
$\lim _{R(T) \to \infty, K\to \infty} (K+1)^2/R(T)^2\approx k^2$ ratio, it is converted into a H Atom-like  spectrum formula,
\begin{eqnarray}
E_K=Ry^s\left(-\frac{1}{(K+1)^2} +\left(\left(a_0^q\right)^\ast \right) ^2 k^2 \right). 
\label{Rlarge}
\end{eqnarray}

The first term in the latter  expression describes the bound states of a quark 
 within the Coulomb-like strong potential,
\begin{equation}
V_C(r)= -\frac{\hbar c \alpha_s N_c}{r},
\label{clpsd_pt}
\end{equation}
while the second term represents the corresponding scattering states.
The $Ry^s$ value from (\ref{curved_spctrm}) now fixes the ionization energy.
This is caused by the collapse of the `curved'  color-electric charge dipole potential down to a flat Coloumbian-type one.
Indeed, from (\ref{curved_dipole}) one reads off that in the small angle approximation,  $\chi=\stackrel{\frown}{r}/R(T)$, i.e., for arc-length values  $\stackrel{\frown}{r}$  small compared to $R(T)$,   
\begin{eqnarray}
\lim_{\chi \to 0}\Gamma_N(\chi)=\alpha_s N_c\left( \cot\chi -\frac{1}{\pi}\right), &\quad& \lim_{\chi\to 0}\Gamma_S(\chi)= -\frac{\alpha_sN_c}{\pi}\,.
\label{Lim1}
\end{eqnarray}
Next, in the same regime we have the following approximation to $\cot\chi$,
\begin{equation}
\lim _{\chi \to 0}\cot \chi\approx \frac{1}{\chi }\,.
\end{equation} 
With these results, the `curved' color-electric charge dipole potential  in (\ref{Gl_6}), when considered  in the large $R(T)$ limit, becomes 
\begin{equation}
-\lim_{R(T)\to \infty}\frac{\hbar c }{R(T)}\alpha_s N_c\cot \chi=
-\frac{\hbar c }{R(T)}\alpha_s N_c \frac{1}{\frac{\stackrel{\frown}{r}}{R(T)}}=
-\hbar c\alpha_s N_c \frac{1}{\stackrel{\frown}{r}}\stackrel{
\stackrel{\frown}{r}\to r}{\longrightarrow } -\frac{\hbar c \alpha_s N_c}{r},
\label{cllps}
\end{equation}
which is Coulombian potential for $\stackrel{\frown}{r} \to r $. Notice, however, that small angles do not necessarily imply small arc lengths, this because the arc lengths are also  growing with the radius. For this reason, the extension of the flattened region around the particle's position can still become significant.
Furthermore, it has been shown in the literature in \cite{Barut}, \cite{aggoun} that also the wave functions of the `curved' problem (admittedly, for the electromagnetic case of the H Atom) collapse to the Coulomb wave functions  in the large compactification radius limit.

\subsection{Effect of the compactification radius on the strong coupling.Rydberg-atom like multi-hadron systems}

In this section we consider multi--quark-anti-quark hadrons, i.e. hadrons of the kind of $2n$ body systems of color-electric charges for $n>1$, the simplest being a tetra-quark.  For this reason, the potential in (\ref{Gl_6}) will be interpreted as the one generated by a  color-electric charge dipole in (\ref{curved_dipole}). Our point is that specifically such systems are those most pre-disposed to color deconfinement.\\

To begin with, in \cite{KTV} we extended in the equation (9.11) the QCD formula for the strong coupling, corresponding to its calculation  to one loop order,  to incorporate the  compactification radius according to
\begin{eqnarray}
\frac{\alpha_s(Q^2)}{4\pi }\approx \frac{1}{\beta_0\ln \left( \frac{Q^2c^2}{\Lambda_{QCD}^2} + \frac{\hbar^2c^2}{R^2\Lambda_{QCD}^2}\right)},\quad \beta_0=11-\frac{2}{3}n_f,
\label{9_11}
\end{eqnarray}
with $n_f$ standing for the number of flavors, this with the aim to explain its walking toward a fixed value in the infrared.
In the present work,  we slightly modify the latter expression towards a more general one, which admits  for a temperature dependence of the compactification radius, and is given by 
\begin{eqnarray}
\frac{\alpha_s(Q^2)}{4\pi }\approx \frac{1}{\beta_0\ln \left( \frac{Q^2c^2}{\Lambda_{QCD}^2} + \sqrt{1-\rho^2}\, \frac{\hbar^2c^2}{R(T)^2\Lambda_{QCD}^2}
+\rho \frac{R(T)^2\Lambda_{QCD}^2}{\hbar^2c^2}
\right)},
\label{9_11_extns}
\end{eqnarray}
being $\rho$ a free parameter, to be adjusted to data.

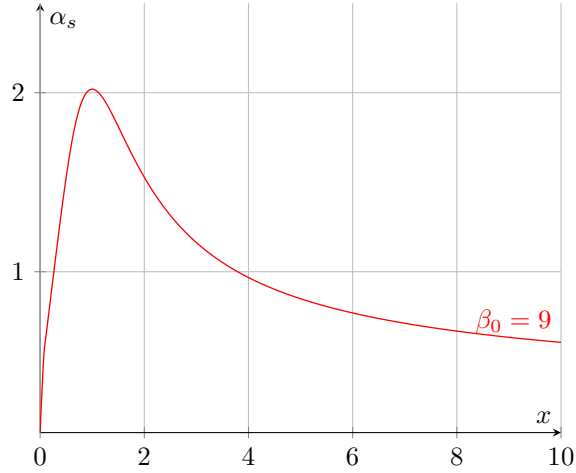
\begin{figure}[H]
  \begin{center}
\begin{tikzpicture}

\begin{axis}[grid=both,
		  xlabel=$x$,ylabel=$\alpha_s$,
          xmax=10,ymax=2.5,
          axis lines=middle,max space between ticks=50pt,
          extra x ticks=0]
\addplot[smooth,black,mark=none,
line width=0.5,domain=0.000001:10,
samples=150,
color=red]  {1.4/ln(x + 1/x)} node[above left] {$\beta_0 =9$};
\end{axis}

\end{tikzpicture}

  \caption{
The dependence of the strong coupling $\alpha_s$ in the infrared regime, 
$Q^2\to 0$ on the compactification radius according to (\ref{9_11_extns}) 
for $\rho=1/\sqrt{2}$, being  $x=\frac{\hbar^2c^2}{R(T)^2\Lambda_{QCD}^2}$. }
\end{center}
  \end{figure}

For  a compactification radius diminishing  with the fall of the temperature, and for increasing values of the transferred squared momentum, the correct behavior of the strong coupling in the ultraviolet (UV), where $Q^2\to \infty$, is reproduced by (\ref{9_11}) as    
\begin{eqnarray}
UV:\qquad \lim _{ R(T)\stackrel{T\to 0}{\longrightarrow} 0}\frac{\alpha_s(Q^2, T)}{4\pi }|_{
Q^2\to \infty}
&\longrightarrow & 0.
\label{9_11_extns_UV}
\end{eqnarray}

Assuming now that the compactification radius increases with the rise of the temperature due, for example, to its linear dependence on $T$,  we find a falling  strong coupling in the infrared (IR) where  $Q^2\to 0$,
\begin{eqnarray}
IR:\qquad \lim _{ R(T)\stackrel{T\to \infty}{\longrightarrow}\infty}\frac{\alpha_s(Q^2, T)}{4\pi }|_{
Q^2\to 0}
&\longrightarrow & 0,
\label{9_11_extns_IR}
\end{eqnarray} (see Fig.~2). 
In this way, at high temperatures (equivalently, large compactification radii), quarks can become asymptotically free at large distances.
This idea of a strong coupling as a function of not only $Q^2$, but also a temperature--dependent compactification radius $R(T)$, as in (\ref{9_11}), is in agreement with the temperature behavior  of 
$\alpha_s(T)$ obtained in  \cite {Steffens}.
If such were to happen, and in combination with the collapse of the color electric charge dipole potential in (\ref{cllps}) to one of the Coulombian type,
at  the approximately flat  surroundings of the heavy $b$ quark position (as depicted in Fig.~3), a hydrogen-like $b{\bar u}$ quark system of the Rydberg-atom type could form.  Placed on a high orbit,  the light quark (an ${\bar u}$ quark in  Fig.~3) can become bound in an extremely weak manner, and suffer either auto-ionization, or being knocked out through an  inelastic scattering collision,  both processes well studied in atomic physics \cite{Ling}, \cite{Garret}. Thus, it is feasible to think of perturbative Abelian approaches in the infrared regime.

\begin{figure}[H]\centering
\begin{tikzpicture}[scale=3]

    \path[draw] (0,0) circle (1);

    \begin{scope}[viewport={\RotationX}{\RotationY}]
            
    \draw[variable=\t, smooth, gray, thick] 
           plot[domain=-90:90] (\ToXYZ{-30}{\t});
    \draw[variable=\t, smooth, gray, thick, dashed] 
           plot[domain=90:110] (\ToXYZ{-30}{\t});
    \draw[variable=\t, smooth, gray, dashed] 
           plot[domain=-90:-120] (\ToXYZ{-30}{\t});                      
            
    \node[circle, fill=red, inner sep=1pt, label={-80:$\overline{u}$}] at (\ToXYZ{-30}{70}) {};
    \node[circle, fill=red, inner sep=1pt, label={280:$d$}] at (\ToXYZ{-30}{-110}) {};        
    \node[circle, fill=blue, inner sep=1pt, label={12:$b$}] at (\ToXYZ{150}{80}) {};    
    \node[circle, fill=blue, inner sep=1pt, label={-120:$\overline{b}$}] at (\ToXYZ{-30}{-80}) {};        
    \draw[gray,very thin,->] (\ToXYZ{0}{90}) -- (0,0,1.35);  
    \draw[gray,very thin]  (\ToXYZ{0}{90}) -- (\ToXYZ{0}{-90});
    \draw[red]  (\ToXYZ{-30}{70}) -- (\ToXYZ{-30}{-110});
    \draw[blue]  (\ToXYZ{150}{80}) -- (\ToXYZ{-30}{-80});     
    \node[circle,fill=black, inner sep=1pt] at (0,0) {};  
        
    \end{scope}

\end{tikzpicture}
\begin{tikzpicture}[scale=1.25]

\draw[] (0,0) -- (1,1.25) -- (5,1.25) -- (4,0) -- cycle;
\draw[very thick, gray] (1,0.5) to[bend left=10] (3.5,0.85);

\draw[dashed, blue] (3.5,0.85) to (3.46,0);
\draw[blue] (3.46,0) to (3.3,-3.75);

\draw[red] (1.27,0) to (3.3,-3.75);
\draw[dashed, red] (1,0.5) to (1.27,0);

\draw[gray, very thin, ->]  (3.3,0.5) to (3.3,1.5);
\draw[dashed, gray, very thin]  (3.3,0.5) to (3.3,0);
\draw[gray, very thin]  (3.3,0) to (3.3,-3.75);

\node[circle, fill=red, inner sep=2pt, label={$\overline{u}$}] at (1,0.5) {};
\node[circle, fill=blue, inner sep=2pt, label={$b$}] at (3.5,0.85) {};
\end{tikzpicture}
\caption{The heavy tetra-quark $ b {\bar b}\, {\bar u}\, d$ on $S^3$. The heavy $b \bar b$ color--electric charge dipole  is considered as the source of the potential in (\ref{tRM_KK}).  The light $\bar u$ quark moves in the upper hemisphere  within the `curved' field, a motion mirrored in the lower hemisphere by the light $d$ quark (left). In the large compactification radius limit, the gradual vanishing of the strong coupling can lead to $(b{\bar u} ) $-- $({\bar b}d)$ dissociation and to a formation of $b\bar u$, and ${\bar b}d$ Rydberg-type meson states,
predisposed to ionization. In this way, we have a quantum mechanical deconfinement mechanism (right).}
\end{figure}
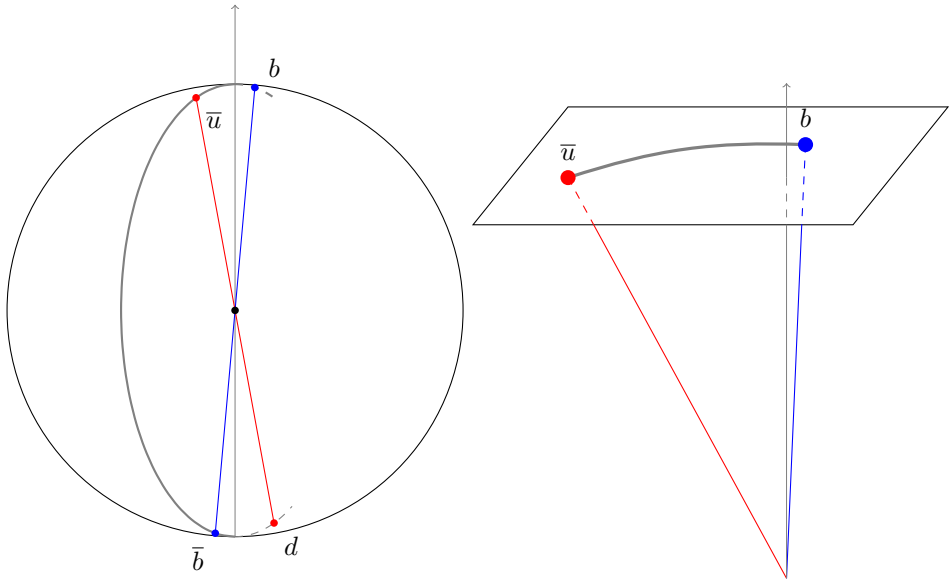

As to the linear dependence between the compactification radius and the temperature, it could  be  parameterized on the basis of dimensional arguments as
\begin{equation}
T=\frac{1}{N_c} \frac{\Lambda_{QCD}^2}{\hbar c}R.
\label{T_T_lin}
\end{equation}
{}For $\Lambda_{QCD}=200$ MeV, and with the result $R=0.58$ fm  according to 
\cite{KC2016}, the temperature, $T_{\mbox{b-st}}$,  of the bound state spectra is 
estimated as
\begin{equation}
T_{\mbox{b-st}}=39.19\, \mbox{MeV}.
\label{Spctrm_Tmpr}
\end{equation}
This admittedly very crude estimate, matches surprisingly well
the one obtained from a QCD--based analysis, as reported  in  \cite{Hands} for $T\sim 40$ MeV. This result offers some support for our hypothesis that a quantum mechanical description of the main aspects of quark deconfinement is possible.

There are some limitations, though. According to (\ref{T_T_lin}) the compactification radius for Hagedorn's (pseudo)critical temperature of $T_H\sim 160$ MeV \footnote{Also confirmed by \cite{AramDavid} for the transition to a Bose-Einstein condensate of a charmonium gas, when described by the partition function of the trigonometric Rosen-Morse potential, considered here in  4/4.3.1 above.}  appears to be around $2$fm, which is admittedly too low a value for the formation of the quarkish Rydberg atom scenario. In this temperature range, no quantum mechanical deconfinement mechanism seems to exist. A reasonable value of $R\sim 10^2$ fm would require, according to (\ref{T_T_lin}), the temperature to rise about $7$ GeV and higher. The Bohr radius at this temperature will also significantly grow due to the decrease of the strong coupling according to (\ref{9_11_extns_IR}).

From  these values onward,  the strong coupling starts walking asymptotically to zero and the 
quantum mechanical mechanism of deconfinement, via the formation of Rydberg-like multi-quark systems and their ionization,  begins to acquire importance.

\section{Conclusions and outlook}
The results reported in the present work can be summarized as follows.
\begin{itemize}
\item  A new family of scalar potentials  could be constructed by the  Master equation (\ref{new_class}) through  conformal metric deformations of the compactified Minkowski spacetime $M^c$, with their  ground states  predicted in (\ref{U-sltn}) by properties of the conformal wave operators in (\ref{fund_thrm}).
\item Three of these potentials, are the exactly solvable  trigonometric Rosen-Morse-, P\"oschl-Teller and Scarf potentials. In this way, we assign a geometric meaning to the known SUSY-QM potentials, which is shared with more general potentials, whose exact ground state wave functions belong to $so(4)$ carrier spaces. 
\item The scheme allows for a  straightforward extension towards hyperbolic topologies of the type $S^1\times H_\pm ^3$ and $S^1\times H^3$, which can be useful to study hyperbolic potentials.

\item The potentials induced by metric deformations considered in Section \ref{sec4}  have been interpreted as mean field potentials, simulating the overall internal gauge field dynamics. They all are suited for the description of the ground states of confined two-body systems on the compactified Minkowski spacetime.
 In this way,  a duality between metric deformations and scalar potentials  has been detected  in dimensions higher than two (in those no gradient terms at all appear upon metric deformation). This could benefit studies performed by means of simulation techniques of quantum confinement phenomena depending on the ground state properties of two-body systems on $M^c$.

\item In  Subsection \ref{sec52} we discussed the `reduced mass' parameter appearing in Schr\"odinger's version of the massless conformal wave equation (\ref{tRM_KK}) on $S^1\times S^3$, which we found to originate in the $\Lambda_{QCD}$ scale parameter of QCD in (\ref{Gl_6_2}), and not from the particle masses as usually assumed. In this way, Schr\"odinger's wave equation in (\ref{Gl_6})  remains  compatible with the masslessness of the  scalar particle coupled to the deformed $S^1\times S^3$ metric according to  (\ref{reduced_dAl}).  

\item In Section \ref{sec5} we considered the effect of the  compactification radius on the color-electric charge dipole potential in (\ref{tRM_undercover}).  It was observed  that, in the  limit of a large compactification radii, of order $R\sim 10^2$ fm (equivalently, temperatures of order $\sim 7$ GeV and higher),  this potential  collapses down to a Coloumbian one in the flattened space around a point on the $S^3$ hyper-sphere, thus leading to a deconfinement mechanism.

\item This process allows for the formation of hydrogen-like quarkish Rydberg-states, known to be easily ionizable. Indeed, the Rydberg quark, occupying  an orbit with a high principal quantum number, behaves approximately as a free particle and can be knocked out in an inelastic scattering event.
\end{itemize}
 
In summary, the model here  presented  allows for a smooth deconfinement transition without loss of conformal symmetry, and without any need for $M^c$  decompactification mechanisms. The compactification radius has been identified as a parameter of the confinement-deconfinement transition, and a quite reasonable spectrum temperature of $T\sim 40$ MeV, has been estimated in (\ref{Spctrm_Tmpr}), a value compatible with estimates by other authors.
Thus, our description of the  confinement-deconfinement phenomena in QCD is mainly a quantum mechanical one, based on some topological properties of $M^c$. In previous works, quark confinement has been considered in terms of Dirac equations on Riemannian  spaces of constant curvature (in \cite{Celso}). Also, more recently,  axiomatic field theories have been shown to generalize to spaces with constant curvatures, the De Sitter and anti-De Sitter spaces being the prime candidates, an idea pursued by \cite{Gazeau} and references therein.

%


%



\begin{thebibliography}{99}

\bibitem{Schottenloher}  M.\ Schottenloher,  
{ A Mathematical introduction to conformal field theory;
 Lect.\ Notes Phys.\  759} (Springer, Berlin Heidelberg, Germany, 2007). 


\bibitem{KTV} M.\ Kirchbach, T.\ Popov, and J.\ A.\  Vallejo, 
Color confinement at the boundary of the conformally compactified $AdS_5$,
{ J. High Energy Phys.} 09 (2021) 171.


\bibitem{KC2016} M.\ Kirchbach and C.\ B.\  Compean, 
Modelling duality between bound and resonant meson spectra by means of free quantum motions on the de Sitter space time $dS_4$,
{ Eur. Phys. J. A } 52  (2016) 210.
Addendum: { Eur. Phys. J. A} 53 (2017) 65.

\bibitem{Ahmed} A.\ Al-Jamel, 
Heavy quarkonia properties from a hard-wall confinement potential model with 
conformal symmetry perturbing effects,
 { Mod.\ Phys.\ Lett.\ A} 34 (2019) 1950307.

\bibitem{AbuShady} M.\ Abu-Shady and  Sh.\ Y.\  Ezz-Alarab, Trigonometric Rosen-Morse potential as the quark-anti-quark interaction for meson properties in the non-relativistic quark model using EAIU, Few Body Syst.60 (2019) 66.


\bibitem{MK_Formfactors} M.\ Kirchbach and C.\ B.\ Compean, Proton's electromagnetic form factors from a non-power confinement  potential, 
{ Nucl.\ Phys.\ A } 980 (2018) 32-50.

\bibitem{Marco} M.\ A.\ Bedolla, Kh.\ Raya, and A.\ Raya, A,
Quark confinement from different dressed gluon propagators, 
{ Few Body Syst.} 64 (2023) 47.

\bibitem{AramDavid} A.\  Bahroz Brzo and D.\  Alvarez-Castillo,
 Thermodynamic properties of the trigonometric Rosen-Morse potential and applications to a quantum gas of mesons,
{ Mod.\ Phys.\ Lett.\  A} 36 (2021) 2150095.

\bibitem{Pirner}  H.\ J.\ Pirner and J.\ Wraldsen, Construction of an effective colour dielectric lattice action from QCD blocking,
{ Nucl. Phys. B} 294 (1987) 905-924.

%
%
%
%
%
%


\bibitem{Brodski}  S.\ J.\ Brodsky,  G.\ de T\'eramond, and M.\ Karliner,
Puzzles in hadron physics and novel Quantum Chromodynamic Phenomena,
 { Ann.\ Rev.\ Nucl.\ Part.\ Sci.\ } 62
( 2012) 1-35.


\bibitem{Deur} A.\ Deur, V.\ Burkert, J.\ P.\  Chen, and W.\  Korsch, 
Determination of the effective strong coupling constant $\alpha_s(Q^2)$ from CLAS spin structure function data,
{ Phys.\ Lett. B} 665 (2008) 349-351.


\bibitem{BirrelDavis}  N.\ D.\ Birrel and P.\ C.\ W.\ Davies, 
 { Quantum fields in curved space\/} (Cambridge University Press, Cambridge, UK, 1982)

\bibitem{4harm} Sh.\ Ahmad Akhoon, A.\ Hussain Sofi, A.\ Maini, and R.\ Ahmad,
Spherical harmonics on four sphere, { Appl.\ Math.\ Phys.\ } 2
( 2014)  157-160.


\bibitem{Marinelius} R.\ Marinelius and B.\ Nilsson, Manifestly conformally covariant field equations and the geometric understanding of mass,
{Phys.\ Rev.\ D } 22 (1980) 830-838.

\bibitem{Brigita} B.\ Alertz, Electrodynamics in Robertson-Walker spacetimes,  
{Ann. Inst. Henri Poincar\'e} 53 (1990) 319-342.


\bibitem{Belitsky} A.\ V.\ Belitsky, A.\ S.\  Gorsky, and G.\ P.\ 
 Korchemski,  Gauge/string duality for conformal string operators, 
{ Nucl.\ Phys.\ B} 667 (2003) 3-54.




\bibitem{Khare} R. De, R. Dutt, and U. Sukhatme,
Mapping of shape invariant potentials under point canonical transformations,
J. Phys. A:Math.Gen. 25 (1992) L843-L850.

\bibitem{Gritsev} V. V. Gritsev, Yu. A. Kurochkin, and V. S. Otchik,
Nonlinear symmetry algebra of the MIC-Kepler problem on the sphere $S^3$. J. Phys. A:Math. Gen. 33 (2000) 4903-4910.

\bibitem{Schwinger} J. Schwinger, A magnetic model of matter, Science 165 (1969) 757-761.


\bibitem{ShiDong} Shishan Dong,  G.\ Y\'a${\tilde {\mathrm n} }$ez-Navarro,
M.\ A.\  Mercado S\'anches,  Guo-Hua Sun,  C.\ Mejia Garcia, and Shi-Hai Dong, 
 Constructions of soluble potentials for the non-relativistic system by means of the Heun functions,
{ Adv.\ High Energy Phys.\ } 2018  (2018) 9824538.


\bibitem{Raposo} A.\ Raposo, H.-J.\   Weber, D.\ E.\  Alvarez-Castillo, and 
M.\   Kirchbach, Romanovski polynomials in selected physics problems,
{\it  C.Eur.J.Phys.} 5 (2007)  253-284.


\bibitem{TQC} C.\ B.\ Compean, and M.\ Kirchbach, 
Trigonometric quark confinement potential of QCD traits,
{Eur.Phys.J.~ A} 33 (2007)  1-4.


\bibitem{Schr41} E.\ Schr\"odinger, A method of determining quantum-mechanical eigenvalues and eigenstates,
{  Proc.\ R.\  Irish Acad.\ A } 46 {1940/ 1941}  9-16.



\bibitem{LT14} M.\ Kirchbach, T.\ Popov, and J.\ A.\ Vallejo,
The Conformal Symmetry-Color Neutrality Connection in Strong Interaction, in 
 Dobrev, V.  (eds) {Lie Theory and its Applications in Physics. LT21:
 Springer Proceedings in Mathematics \& Statistics, vol.\ 396}
 (Springer, Singapore, 2021) pp. 361-369.


\bibitem{PRD10} M.\ Kirchbach and C.\ B.\  Compean, Conformal symmetry and light flavor baryon spectra, {Phys. Rev. D} 82 (2010)  034008.
 
\bibitem{Afonin} S.\ S.\  Afonin,  Hydrogen like classification for light non-strange mesons, { Int.\ J.\ Mod.\ Phys.\ A } 23  (2008),  4205-4217.

\bibitem{Afonin3} S.\ S.\ Afonin,  Light meson spectrum and classical symmetries of QCD, { Eur.\ Phys.\ J.\ A} 29  (2006) 327-335.

%
\bibitem{Valcarse} P.\  Gonzalez, Long-distance behavior of the quark-antiquark  static potential. Applications to light quark mesons and heavy quarkonia,
{ Phys.\ Rev.\ D } 80  (2009)  054010.


\bibitem{Barut} A.\ O.\ Barut and R.\  Wilson, On the dynamical group of the Kepler problem in curved space of constant curvature,  Phys.\ Lett.\ A 110  (1985)   351-354.

 \bibitem{Hands} S.\  Hands, T.\ J.\  Hallowood, J.\ C.\  Myers,  
QCD with chemical potential in a small hyperspherical box,
{J.\ High Energy Phys.\ } 1007 (2010) 086.

\bibitem{aggoun} L.\ Aggoun, N.\ Bounouioua, F.\ Benamira, and L.\  Guechi,
Path integral solution for the Coulomb potential in curved space of constant positive curvature, { Int.\ J.\ Theor.\ Phys.\ } 55 (2016)   2553-2667.
   
\bibitem{Ling} X.\ Ling, M.\ T.\  Frey, K.\ A.\  Smith, and F.\ B.\  Dunning,
Inelastic electron-dipole molecule scattering at sub-milli-electron colt energies:HF and $NH_3$,
{ Phys.\ Rev.\ A } 48  (1993)  1252-1256.

\bibitem{Garret}  W.\ R.\ Garrett,   Low energy electron scattering by polar molecules, {  Mol. Phys.} 24  ( 1972 )  465-487.



\bibitem{Steffens} Steffens, F.M. The temperature dependence of the QCD Running coupling, {\it Braz.J.Phys.} (2006) 36, 582-585.

\bibitem{Celso}  C.\ Barros Jr., Quark confinement and curved space.
{Eur. Phys. J.\ C}  45 (2006) 421-425.


\bibitem{Gazeau}  M.\ V.\ Takook,  Axiomatic de Sitter quantum Yang-Mills theory with color confinement and mass gap, { Eur.\ Phys.\ Lett.\ } 141 (2023) 22003.

\end{thebibliography}
\end{document}